\documentclass[%
 reprint,
 amsmath,amssymb,
 aps,
]{revtex4-1}

\usepackage{graphicx}
\usepackage{dcolumn}
\usepackage{bm}
\usepackage{longtable}
\usepackage{enumerate}
\usepackage{subfigure}
\usepackage{multirow}
\usepackage{cancel}
\usepackage{tabularx}
\usepackage{amssymb}
\usepackage{balance} 
\usepackage{rotating}
\usepackage{url}
\usepackage{amsmath,amssymb}
\usepackage{longtable}
\usepackage{tablefootnote}
\usepackage[para]{threeparttable}
\usepackage{tablefootnote}
\usepackage{color} 
\usepackage{bm}
\usepackage{ulem}

\newcommand{\Lower}[1]{\smash{\lower 1.5ex \hbox{#1}}}

\newcommand{\down}{\sout{$\downarrow$}}
\newcommand{\up}{\sout{$\uparrow$}}
\newcommand{\zero}{\sout{\phantom{$\downarrow \negthickspace \uparrow$}}}
\newcommand{\double}{\sout{$\downarrow \negthickspace \uparrow$}}
\newcommand{\ib}{\bar{i}}
\newcommand{\jb}{\bar{j}}

\DeclareMathOperator{\Tr}{Tr}

\begin{document}

\preprint{}
\title{Orbital entanglement in quantum chemistry}
\author{Katharina Boguslawski}
\email{k.boguslawski@fizyka.umk.pl}
\author{Pawe{\l} Tecmer}%
\email{ptecmer@fizyka.umk.pl}
\affiliation{%
 Department of Chemistry and Chemical Biology, McMaster University, Hamilton, 1280 Main Street West, L8S 4M1, Canada \\
}%
\vspace{-2cm}


\date{\today}

\begin{abstract}
The basic concepts of orbital entanglement and its application to chemistry are briefly reviewed. The calculation of orbital entanglement measures from correlated wavefunctions is discussed in terms of reduced $n$-particle density matrices. Possible simplifications in their evaluation are highlighted in case of seniority-zero wavefunctions. Specifically, orbital entanglement allows us to dissect electron correlation effects in its strong and weak contributions, to determine bond orders, to assess the quality and stability of active space calculations, to monitor chemical reactions, and to identify points along the reaction coordinate where electronic wavefunctions change drastically. Thus, orbital entanglement represents a useful and intuitive tool to interpret complex electronic wavefunctions and to facilitate a qualitative understanding of electronic structure and how it changes in chemical processes.

\end{abstract}

\pacs{Valid PACS appear here}
\maketitle


\section{Introduction\\} 
The entangled nature of quantum mechanics has been exploited in many areas of physics and computer science over the past two decades and has led to many important discoveries in quantum cryptography, quantum teleportation and quantum computing.~\cite{Vedral_2002,Horodecki-rev-2009,quantum-comput-2014,quantum-topology-2014}
The phenomena of quantum entanglement assumes the existence of global states of a composite system that, as a result of ``spooky'' interactions, cannot be written as a product of states of the individual subsystems.~\cite{EPR-1935,Schrodinger-entanglement-1935} That is, quantum entanglement implies that even when one has the most complete possible knowledge of the total system, one may remain wholly ignorant about the state of its individual parts.~\cite{Schrodinger-entanglement-1935} To this end, let $|\Psi \rangle$ be the total wavefunction of a pure quantum state. Using the superposition principle, the total state of the system can be written as
\begin{equation}\label{eqn:superposition}
|\Psi \rangle = \sum_{\bm{i}_1, \ldots , \bm{i}_k} c_{\bm{i}_1, \ldots ,\bm{i}_k}|\bm{i}_1\rangle \otimes |\bm{i}_2 \rangle \otimes \ldots \otimes |\bm{i}_k\rangle, 
\end{equation}
where $|\bm{i}_j \rangle$ is the basis of local Hilbert space $\mathcal{H}_j$, $\{c_{\bm{i}_1, \ldots ,\bm{i}_k}\}$ are complex numbers, and $\mathcal{H} = \bigotimes_{l=1}^k \mathcal{H}_l$ is the total Hilbert space. If the quantum state of eq.~\eqref{eqn:superposition} is entangled, it cannot be written as a product of states of the individual subsystems, $| \Psi \rangle \neq | \psi_1 \rangle \otimes |\psi_2 \rangle \otimes \ldots \otimes |\psi_k\rangle$.
As a consequence, a bipartite composite system $AB$ with wavefunction $|\Psi^{AB}\rangle$ cannot always be represented as a product state, $|\Psi^{AB}\rangle=|\psi^{A}\rangle \otimes |\psi^{B}\rangle$, but only as a series of tensor products of basis states of the individual subsystems,
\begin{equation}\label{eqn:entangled-state}
|\Psi^{AB}\rangle = \sum_{i,j}c_{i,j}|\psi_i^A\rangle \otimes|\psi_j^B\rangle.  
\end{equation}
Quantum states described by the above equation are called entangled states, while quantum states that are not entangled (\textit{i.e.}, states that can be written as product states) are called \textit{separable}. In contrast to pure states, entanglement of a mixed state $\rho^{\rm mixed}=\sum_i p_i |\Psi_i \rangle\langle \Psi_i|$ is not defined using the decomposition into product states. A mixed state of $k$ subsystems is entangled if it cannot be written as a convex combination of product states,~\cite{Horodecki-rev-2009}
\begin{equation}\label{eqn:mixed-state}
\rho^{\rm mixed} = \sum_{i} p_{i} \rho_{1,i} \otimes \ldots \otimes \rho_{k,i}.  
\end{equation}

Compared to physics, the utility of quantum entanglement in electronic structure theory has been realized only recently.~\cite{Ziesche_1995,Nagy_1996,Nalewajski_2000,legeza_dbss,Rissler_2006,Juhasz_2006,Luzanov_2007,Alcoba_2010,orbitalordering,entanglement_letter,gergo-2014}
In this perspective, we will discuss how quantum entanglement can be used to measure orbital interactions and how it can provide a different perspective on well-established concepts in quantum chemistry.

\section{Entropy Measures\\}
A quantitative measure of entanglement is provided by the von Neumann entropy 
\begin{equation}\label{eq:vN}
S =-\Tr(\rho \ln \rho),
\end{equation}
where $\rho$ is the density matrix of the quantum system. For a bipartite system divided into two parts $A$ and $B$, the entanglement between $A$ and $B$ is defined in terms of the density matrices of each subsystem. The entanglement entropy $S_{A|B}$ quantifies the interaction (in terms of informational exchange) between $A$ and $B$. For pure states, we have
\begin{equation}\label{eq:ee}
S_{A|B} =-\Tr(\rho_A \ln \rho_A) =-\Tr(\rho_B \ln \rho_B)
\end{equation}
with the reduced density matrix (RDM) of each part,
\begin{equation}\label{eq:rdm}
\rho_A =\rm {Tr}_B |\Psi\rangle \langle \Psi| \quad {\rm and \quad} \rho_B =\rm {Tr}_A |\Psi\rangle \langle \Psi|,
\end{equation}
where $\rho_A$ ($\rho_B$) is calculated by tracing out the states on subsystem $B$ ($A$). Thus, the eigenvalue spectrum of $\rho_A$ (which is equivalent to the eigenvalue spectrum of $\rho_B$) characterizes the entanglement between subsystems $A$ and $B$. 

\begin{figure*}[t]
\caption{\label{fig:partitioning}
Possible partitionings of the orbitals in different subsystems $A$ and $B$. The two subsystems are color-coded. The orbitals are grouped into a left (l), middle (m), and right (r) block with states $\bm{n}_l,\bm{n}_m,$ and $\bm{n}_r$. On the right hand side, the wavefunction of the composite system $AB$ is displayed. The orbital blocks can be reordered so that the orbital bath forms a continuous block. The wavefunction can then be written as $|\Psi^{AB}\rangle=\sum_{\bm{i},\bm{n}} \tilde{c}_{\bm{n},\bm{i}} |\bm{n}^A\rangle \otimes |\bm{i}^B\rangle$, introducing a proper phase factor.
}
\vspace*{-0.5cm}
\begin{center}
\includegraphics[width=0.99\textwidth,keepaspectratio=true]{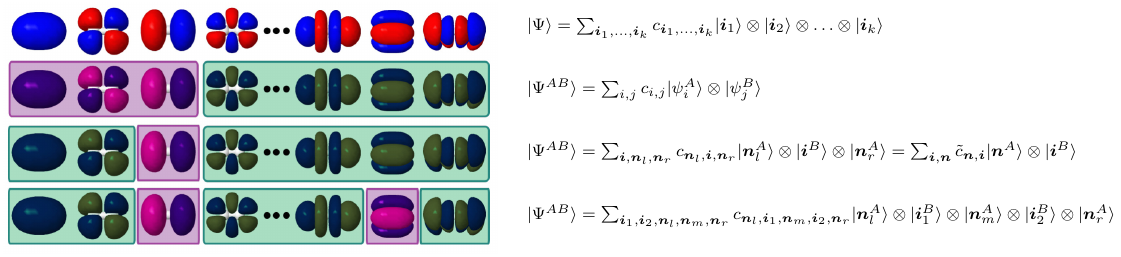}
\end{center}
\vspace*{-0.5cm}
\end{figure*}

So far, we have made no assumptions about the choice of the composite system and its parts. In quantum chemistry, the interaction of orbitals is commonly used to understand chemical processes. Examples are molecular orbital diagrams, frontier orbital theory, and ligand field theory.  Since a unique quantitative measure of orbital interactions does not exist, conventional models of interacting orbitals are based on qualitative arguments. To provide an alternative perspective, the concept of entanglement can be used to quantify the interaction of orbitals. For that purpose, consider a set of $k$ orbitals that span the $N$-particle Hilbert space. Now, we can partition the Hilbert space into (any) two parts $A$ and $B$ where possible partitionings are displayed in Figure~\ref{fig:partitioning}. If the Hilbert space is divided into one orbital (with local basis states $\{|\bm{i}\rangle\}$) and the space spanned by all remaining orbitals ($\{ |\bm{n} \rangle\}$), the entanglement entropy quantifies the interaction of this particular orbital (in terms of exchange of information) and the orbital ``bath''. To emphasize the partitioning of the $N$-particle Hilbert space, the entanglement entropy of one orbital will be denoted as single-orbital entropy (or one-orbital entropy).

Using eq.~\eqref{eq:rdm}, the single-orbital entropy $s(1)_i$ is determined from the eigenvalues $\omega_{\alpha,i}$ of the one-orbital
RDM $\rho_{i,i'}^{(1)}$ of a given orbital $i$
\begin{equation}\label{eq:rho1}
\rho_{i,i'}^{(1)}= \sum_{\bm{n}} \langle \bm{n}| \langle i| \Psi\rangle \langle \Psi| i' \rangle |\bm{n} \rangle .
\end{equation}
Substituting the above equation into eq.~\eqref{eq:vN}, we obtain
\begin{equation}\label{eq:s_1}
s(1)_i=-\sum_{\alpha=1}^4 \omega_{\alpha,i}\ln \omega_{\alpha,i}.
\end{equation}
Since the one-orbital RDM is determined from an $N$-particle RDM by tracing out all other orbital-degrees of freedom except those of orbital $i$, this leads to a RDM whose dimension is equal to the dimension of the one-orbital Fock space. In the case of spatial orbitals, four different states (occupations of orbitals) exist, which can be empty $|$\zero$\rangle$, occupied with an $\alpha$-(spin-up) $|$\up$\rangle$ or a $\beta$-(spin-down) $|$\down$\rangle$ electron, or doubly occupied with two electrons of paired spin  $|$\double$\rangle$.

The sum of single-orbital entropies defines the total quantum information encoded in the
system,
\begin{equation}\label{eq:I_tot}
I_{\rm tot} = \sum_i s(1)_i.
\end{equation}
However, other possibilities to partition the system into two parts than one orbital and an orbital bath exist. If the Hilbert space is divided into a subspace spanned by two orbitals and the space spanned by the remaining $k-2$ orbitals, the entanglement entropy quantifies the interaction between one orbital pair and the orbital bath. This two-orbital entropy $s(2)_{i,j}$ is determined from two-orbital RDM, which can be calculated similar to eq.~\eqref{eq:rho1}, but $\{|\bm{i}\rangle\}$ being the basis states of a two-orbital Fock space
(with 16 possible states for spatial orbitals: $|$\zero \,\zero$\rangle$, $|$\zero \,\down$\rangle$,  $|$\down \,\zero$\rangle$, $|$\zero \,\up$\rangle$, ~\dots, $|$\double \,\double$\rangle$).
The two-orbital analogue of Eq~\eqref{eq:s_1} is
\begin{equation}\label{eq:s_ij}
s(2)_{i,j} =-\sum_{\alpha=1}^{16} \omega_{\alpha, i, j} \ln \omega_{\alpha, i, j},
\end{equation}
where $\omega_{\alpha, i, j}$ are the eigenvalues of the two-orbital RDM.

The total amount of correlation between any pair of orbitals $(i,j)$ can be evaluated from the (orbital-pair) mutual information. Specifically, the mutual information allows us to measure the total amount of information one system (here, orbital $i$) has about another system (here, orbital $j$), including all types of correlation (classic and quantum).~\cite{Groisman2005,Wolf2008,gergo-2014} The orbital-pair mutual information~\cite{Rissler_2006} is calculated using the single- and two-orbital entropy and thus requires the one- and two-orbital RDMs,
\begin{equation}\label{eq:I_ij}
I_{i|j} = \frac{1}{2} \big(s(2)_{i,j} - s(1)_{i} - s(1)_{j} \big) \big(1 - \delta_{ij}\big), 
\end{equation}
where $\delta_{ij}$ is the Kronecker delta.

It is obvious from eqs.~\eqref{eq:ee} and \eqref{eq:rdm} that a correlated wavefunction is required to have non-zero orbital entanglement and correlation. In the case of an uncorrelated wavefunction, \textit{e.g.}, a single Slater determinant, the (orbital) entanglement entropy is zero.~\cite{Rissler_2006} Furthermore, we should emphasize that reliable orbital entanglement can only be obtained from (correlated) wavefunctions that provide a good approximation to the full-configuration interaction (FCI) solution. 

In this perspective, we focus solely on orbital entanglement (eq.~\eqref{eq:s_1}) and correlation (eq.~\eqref{eq:I_ij}) where the $N$-particle Hilbert space is partitioned into orbital subspaces. Different entanglement measures that are based on different partitioning schemes and the general definition of the von Neumann entropy eq.~\eqref{eq:vN} have also been introduced in quantum chemistry by other authors.~\cite{Luzanov_2005,Juhasz_2006,Luzanov_2007,kurashige2013}

\section{Entropy Measures from Density Matrices\\}
In contrast to $n$-particle RDMs that are defined for a constant number of particles, the one- and two-orbital RDMs contain information from the 1-, 2-, 3-, and 4-particle RDM and are thus defined for a variable number of particles. Specifically, the one-orbital RDM elements can be determined from a subset of elements of the spin-dependent 1-RDM and 2-RDM, while the two-orbital RDM requires in addition specific elements of the 3- and 4-RDM. Furthermore, if $s_z$ is a good quantum number, the one-orbital RDM is diagonal, whereas the two-orbital RDM is block-diagonal and contains only non-zero elements for two-orbital states that preserve the particle number and $s_z$.
In terms of the spin-dependent 1- and 2-RDMs, $\gamma_j^i=\langle a_i^\dagger a_j \rangle$ and $\Gamma_{kl}^{ij}=\langle a^\dagger_ia^\dagger_ja_la_k\rangle$, respectively, $\rho_i^{(1)}$ is given as
\begin{equation*}
\rho_i^{(1)} =
\begin{pmatrix}
1-\gamma_{i}^{i}-\gamma_{\bar{i}}^{\bar{i}}+\Gamma_{i\bar{i}}^{i\bar{i}} & 0 & 0 & 0 \\
0 & \gamma_{{i}}^{{i}}-\Gamma_{i\bar{i}}^{i\bar{i}} & 0 & 0 \\
0 & 0 & \gamma_{\bar{i}}^{\bar{i}}-\Gamma_{i\bar{i}}^{i\bar{i}} & 0\\
0 & 0 & 0 & \Gamma_{i\bar{i}}^{i\bar{i}} 
\end{pmatrix},
\end{equation*}
where the indices $i$ and $\bar{i}$ indicate $\alpha$- and $\beta$-electrons and the order of the one-orbital states is $\{$\zero,\up,\down,\double$\}$. The elements of $\rho_{i,j}^{(2)}$ are summarized in Table~\ref{tbl:rho2}. Note that $\rho_{i,j}^{(2)}$ requires only some diagonal elements of the 3- and 4-RDM, as well as a few off-diagonal elements of the 1-,2-, and 3-RDM.
\begin{table*}
\begin{threeparttable}[t]
\centering\tiny
\caption{$\rho_{i,j}^{(2)}$ expressed in terms of $n$-particle RDMs. ${}^3\Gamma_{lmn}^{ijk}=\langle a^\dagger_ia^\dagger_ja^\dagger_ka_na_ma_l\rangle$ and ${}^4\Gamma_{mnop}^{ijkl}=\langle a^\dagger_ia^\dagger_ja^\dagger_ka^\dagger_la_pa_oa_na_m\rangle$.}\label{tbl:rho2}
\begin{tabular}{c|cccccccccccccccc}
\hline
                & \zero\,\zero & \zero\,\up & \up\,\zero & \zero\,\down & \down\,\zero & \up\,\up & \down\,\down & \zero\,\double & \up\,\down & \down\,\up & \double\,\zero & \up\,\double & \double\,\up & \down\,\double & \double\,\down & \double\,\double \\[0.05in] \hline
 \zero\,\zero     & (1,1) & 0 & 0 & 0 & 0 & 0 & 0 & 0 & 0 & 0 & 0 & 0 & 0 & 0 & 0 & 0 \\[0.05in]
 \zero\,\up       & 0 & (2,2) & (2,3) & 0 & 0 & 0 & 0 & 0 & 0 & 0 & 0 & 0 & 0 & 0 & 0 & 0 \\[0.05in]
 \up\,\zero       & 0 & (3,2) & (3,3) & 0 & 0 & 0 & 0 & 0 & 0 & 0 & 0 & 0 & 0 & 0 & 0 & 0 \\[0.05in]
 \zero\,\down     & 0 & 0 & 0 & (4,4) & (4,5) & 0 & 0 & 0 & 0 & 0 & 0 & 0 & 0 & 0 & 0 & 0 \\[0.05in]
 \down\,\zero     & 0 & 0 & 0 & (5,4) & (5,5) & 0 & 0 & 0 & 0 & 0 & 0 & 0 & 0 & 0 & 0 & 0 \\[0.05in]
 \up\,\up         & 0 & 0 & 0 & 0 & 0 & (6,6) & 0 & 0 & 0 & 0 & 0 & 0 & 0 & 0 & 0 & 0 \\[0.05in]
 \down\,\down     & 0 & 0 & 0 & 0 & 0 & 0 & (7,7) & 0 & 0 & 0 & 0 & 0 & 0 & 0 & 0 & 0 \\[0.05in]
 \zero\,\double   & 0 & 0 & 0 & 0 & 0 & 0 & 0 & (8,8) & (8,9) & (8,10) & (8,11) & 0 & 0 & 0 & 0 & 0 \\[0.05in]
 \up\,\down       & 0 & 0 & 0 & 0 & 0 & 0 & 0 & (9,8) & (9,9) & (9,10) & (9,11) & 0 & 0 & 0 & 0 & 0 \\[0.05in]
 \down\,\up       & 0 & 0 & 0 & 0 & 0 & 0 & 0 & (10,8) & (10,9) & (10,10) & (10,11) & 0 & 0 & 0 & 0 & 0 \\[0.05in]
 \double\,\zero   & 0 & 0 & 0 & 0 & 0 & 0 & 0 & (11,8) & (11,9) & (11,10) & (11,11) & 0 & 0 & 0 & 0 & 0 \\[0.05in]
 \up\,\double     & 0 & 0 & 0 & 0 & 0 & 0 & 0 & 0 & 0 & 0 & 0 & (12,12) & (12,13) & 0 & 0 & 0 \\[0.05in]
 \double\,\up     & 0 & 0 & 0 & 0 & 0 & 0 & 0 & 0 & 0 & 0 & 0 & (13,12) & (13,13) & 0 & 0 & 0 \\[0.05in]
 \down\,\double   & 0 & 0 & 0 & 0 & 0 & 0 & 0 & 0 & 0 & 0 & 0 & 0 & 0 & (14,14) & (14,15) & 0 \\[0.05in]
 \double\,\down   & 0 & 0 & 0 & 0 & 0 & 0 & 0 & 0 & 0 & 0 & 0 & 0 & 0 & (15,14) & (15,15) & 0 \\[0.05in]
 \double\,\double & 0 & 0 & 0 & 0 & 0 & 0 & 0 & 0 & 0 & 0 & 0 & 0 & 0 & 0 & 0 & (16,16) \\[0.05in]
\hline
\hline
\end{tabular}
\begin{tablenotes}\footnotesize
\item $(2,3) = (3,2)^\dagger = \gamma_{i}^{j}-\Gamma_{i \ib}^{j \ib}-\Gamma_{i \jb}^{j \jb}+{}^3\Gamma_{i \ib \jb}^{j \ib \jb}$
\item $(4,5)=(5,4)^\dagger = \gamma_{\ib}^{\jb}-\Gamma_{i \ib}^{i \jb}-\Gamma_{j \ib}^{j \jb} + {}^3\Gamma_{ij\ib}^{ij\jb}$
\item $(6,6)= \Gamma_{ij}^{ij}-{}^3\Gamma_{i\ib j}^{i\ib j}-{}^3\Gamma_{ij\jb}^{ij\jb}+{}^4\Gamma_{i\ib j \jb}^{i\ib j \jb}$
\item $(7,7)= \Gamma_{\ib \jb}^{\ib \jb}-{}^3\Gamma_{i \ib \jb}^{i\ib \jb}-{}^3\Gamma_{\ib j\jb}^{\ib j\jb}+{}^4\Gamma_{i\ib j \jb}^{i\ib j \jb}$
\item $(8,8)= \Gamma_{j\jb}^{j\jb}-{}^3\Gamma_{i j\jb}^{ij\jb}-{}^3\Gamma_{\ib j\jb}^{\ib j\jb}+{}^3\Gamma_{i\ib j \jb}^{i\ib j \jb}$
\item $(8,9)=(9,8)^\dagger = \Gamma_{i\jb}^{j\jb}-{}^3\Gamma_{i\jb \ib}^{j \jb \ib}$
\item $(8,10)=(10,8)^\dagger = -\Gamma_{j\ib}^{j\jb}+{}^3\Gamma_{ij \ib}^{i j \jb}$
\item $(8,11)=(11,8)^\dagger = \Gamma_{i\ib}^{j\jb}$
\item $(9,9)= \Gamma_{i\jb}^{i\jb}-{}^3\Gamma_{i\ib \jb}^{i\ib \jb}-{}^3\Gamma_{i j \jb}^{i j \jb}+{}^4\Gamma_{i\ib j\jb}^{i\ib j\jb}$
\item $(9,10)=(10,9)^\dagger = -\Gamma_{j\ib}^{i\jb}$
\item $(9,11)=(11,9)^\dagger = \Gamma_{i\ib}^{i\jb}-{}^3\Gamma_{ij\ib}^{ij\jb}$
\item $(10,10)= \Gamma_{\ib j}^{\ib j}-{}^3\Gamma_{i\ib j}^{i\ib j}-{}^3\Gamma_{\ib j \jb}^{\ib j \jb}+{}^4\Gamma_{i\ib j\jb}^{i\ib j\jb}$
\item $(10,11)=(11,10)^\dagger = -\Gamma_{i\ib}^{j\ib}+{}^3\Gamma_{i\ib \jb}^{j\ib \jb}$
\item $(11,11)= \Gamma_{i\ib}^{i\ib}-{}^3\Gamma_{i\ib j}^{i \ib j}-{}^3\Gamma_{i \ib \jb}^{i \ib \jb}+{}^4\Gamma_{i\ib j\jb}^{i\ib j\jb}$
\item $(12,12)= {}^3\Gamma_{ij\jb}^{ij\jb}-{}^4\Gamma_{i\ib j\jb}^{i\ib j\jb}$
\item $(12,13)= (13,12)^\dagger = -{}^3\Gamma_{ij\ib}^{ij\jb}$
\item $(13,13)= {}^3\Gamma_{i\ib j}^{i\ib j}-{}^4\Gamma_{i\ib j\jb}^{i\ib j\jb}$
\item $(14,14)= {}^3\Gamma_{\ib j \jb}^{\ib j \jb}-{}^4\Gamma_{i\ib j\jb}^{i\ib j\jb}$
\item $(14,15)= (15,14)^\dagger = -{}^3\Gamma_{\ib i \jb}^{\ib j \jb}$
\item $(15,15)= {}^3\Gamma_{i\ib \jb}^{i\ib \jb}-{}^4\Gamma_{i\ib j\jb}^{i\ib j\jb}$
\item $(16,16)= {}^4\Gamma_{i\ib j\jb}^{i\ib j\jb}$
\item $(1,1) = 1-\gamma_{i}^{i}-\gamma_{\ib}^{\ib}-\gamma_{j}^j-\gamma_{\jb}^{\jb}+\Gamma_{i\ib}^{i\ib}+\Gamma_{j\jb}^{j\jb}+\Gamma_{ij}^{ij}+\Gamma_{i\jb}^{i\jb}+\Gamma_{\ib j}^{\ib j}+\Gamma_{\ib\jb}^{\ib\jb}-{}^3\Gamma_{ij\jb}^{ij\jb}-{}^3\Gamma_{\ib j\jb}^{\ib j\jb}-{}^3\Gamma_{i\ib j}^{i \ib j}-{}^3\Gamma_{i\ib\jb}^{i\ib\jb}+{}^4\Gamma_{i\ib j\jb}^{i \ib j\jb}$
\item $(2,2) = \gamma_{j}^{j}-\Gamma_{ij}^{ij}-\Gamma_{\ib j}^{\ib j}-\Gamma_{j \jb}^{j \jb}+{}^3\Gamma_{i\jb j}^{i\jb j}+{}^3\Gamma_{i \ib j}^{i \ib j}+{}^3\Gamma_{\ib j\jb}^{\ib j\jb}-{}^4\Gamma_{i \ib j \jb}^{i \ib j \jb}$\\
\item $(3,3)= \gamma_i^i-\Gamma_{i \ib}^{i \ib} -\Gamma_{ij}^{ij} - \Gamma_{i\jb}^{i\jb}+{}^3\Gamma_{ij\jb}^{ij\jb}+{}^3\Gamma_{i\ib j}^{i\ib j} +{}^3\Gamma_{i\ib \jb}^{i\ib \jb} -{}^4\Gamma_{i\ib j\jb}^{i \ib j \jb}$
\item $(4,4)=\gamma_{\jb}^{\jb}-\Gamma_{i\jb}^{i \jb}-\Gamma_{\ib \jb}^{\ib \jb}-\Gamma_{j\jb}^{j\jb}+{}^3\Gamma_{i \ib \jb}^{i \ib \jb}+{}^3\Gamma_{ij\jb}^{ij\jb}+{}^3\Gamma_{\ib j \jb}^{\ib j \jb}-{}^4\Gamma_{i\ib j\jb}^{i\ib j \jb}$
\item $(5,5)=\gamma_{\ib}^{\ib}-\Gamma_{\ib j}^{\ib j}-\Gamma_{\ib \jb}^{\ib \jb}-\Gamma_{i\ib}^{i\ib}+{}^3\Gamma_{\ib j \jb}^{\ib j \jb}+{}^3\Gamma_{i\ib j}^{i \ib j}+{}^3\Gamma_{i \ib \jb}^{i \ib \jb}-{}^4\Gamma_{i\ib j\jb}^{i\ib j \jb}$
\end{tablenotes}
\end{threeparttable}
\end{table*}

As presented in Refs.~\citenum{Rissler_2006,entanglement_bonding_2013}, $\rho_i^{(1)}$ and $\rho_{i,j}^{(2)}$ can be expressed in terms of number, creation, and annihilation operators. In the case of the density matrix renormalization group, $\rho_i^{(1)}$ and $\rho_{i,j}^{(2)}$ are evaluated using transition operators. We refer the reader to Ref.~\citenum{entanglement_bonding_2013} for more details.

\subsection{Entropy Measures from Seniority-Zero Wavefunctions}
If the electronic wavefunction is restricted to the seniority-zero sector, \textit{i.e.}, a CI-expansion with pair-excited Slater determinants only, $\rho_i^{(1)}$ and $\rho_{i,j}^{(2)}$ have a particular simple form. Due to the absence of singly-occupied orbitals, $\rho_i^{(1)}$ reduces to a $2\times2$ matrix, while $\rho_{i,j}^{(2)}$ becomes a $4\times4$ matrix. Using the relations $\gamma_{p}^{p}=\gamma_{\bar{p}}^{\bar{p}}=\Gamma_{p\bar{p}}^{p\bar{p}}$ and $\Gamma_{p\bar{q}}^{p\bar{q}}={}^4\Gamma_{p\bar{p}q\bar{q}}^{p\bar{p}q\bar{q}}$ that are valid for seniority-zero wavefunctions,~\cite{Weinhold1967a} we have
\begin{equation*}
\rho_i^{(1)} =
\begin{pmatrix}
1-\gamma_{i}^{i} & 0 \\
0 & \gamma_{i}^{i} 
\end{pmatrix}
\end{equation*}
for the seniority-zero one-orbital RDM expressed in the basis $\{$\zero,\double$\}$, and
\begin{equation*}
\rho_{i,j}^{(2)} =
\begin{pmatrix}
1-\gamma_{i}^{i} -\gamma_{j}^{j}+\Gamma_{i\bar{j}}^{i\bar{j}} & 0 & 0 & 0 \\
0 & \gamma_{i}^{i}-\Gamma_{i\bar{j}}^{i\bar{j}} & \Gamma^{i\bar{i}}_{j\bar{j}} & 0\\
0 & \Gamma^{j\bar{j}}_{i\bar{i}} & \gamma_{j}^{j}-\Gamma_{j\bar{i}}^{j\bar{i}} & 0\\
0 & 0 & 0 &\Gamma_{i\bar{j}}^{i\bar{j}} 
\end{pmatrix}
\end{equation*}
for the seniority-zero two-orbital RDM expressed in the basis  $\{$\zero\,\zero,\double\,\zero,\zero\,\double,\double\,\double$\}$. We should note that, for a seniority-zero wavefunction, the maximum value for the entanglement entropy of orbital $i$ is $\ln 2$.

\subsection{Orbital entanglement for an Interpretive Picture\\}
In the following, we briefly discuss some examples where orbital entanglement provides a complementary picture for the interpretation of complex electronic wavefunctions. 

\subsection{Electron Correlation Effects}
Entropic measures based on Eq.~\eqref{eq:s_1} and Eq.~\eqref{eq:I_ij} can be used to dissect electron correlation effects into different contributions, provided electronic wavefunctions can be determined accurately.~\cite{entanglement_letter}
Computational studies revealed that strongly-correlated orbitals with large values of s$(1)_i$ and $I_{i|j}$ are important for nondynamic/static (strong) electron correlation effects. If strongly correlated orbitals are encountered, the system bears a significant multi-reference character.
Conversely, weakly entangled orbitals indicate the dominance of dynamic (weak) electron correlation effects that can be effectively described by single-reference methods, like, for instance, those based on coupled cluster (with a low-order cluster operator) or perturbation theories.
Table~\ref{tbl:corr} maps the strength of orbital interactions onto a certain type of correlation effects. It is important
to stress that there exists no rigorous distinction between different types of electron correlation effects and the values tabulated in Table~\ref{tbl:corr} should be interpreted qualitatively, not quantitatively.
\begin{table}[h]
\centering
\caption{Relation between the strength of entanglement and electron correlation
effects. $s(1)_i$ and $I_{i|j}$ denote the single orbital entropy and orbital-pair mutual information, respectively.}\label{tbl:corr}
\begin{tabular}{|c|c|c|}
\hline
Correlation effects & $s(1)_i$ & $I_{i|j}$ \\
\hline
non-dynamic & $> 0.5$     & $\approx10^{-1}$ \\
static      &$0.5-0.1$    & $\approx10^{-2}$ \\
dynamic     &$< 0.1$      & $\approx10^{-3}$ \\
dispersion  & $\approx 0$ & $\approx10^{-4} - 10^{-5}$\\
\hline
\end{tabular}
\end{table}

Computational studies of electron correlation effects using the single-orbital entropy and orbital-pair mutual
information have been performed for a number of molecules including diatomics, transition metal complexes, and actinide compounds.~\cite{orbitalordering,entanglement_letter,entanglement_bonding_2013,CUO_DMRG,Knecht2014,PCCP_bonding,Corinne-2014}
Specifically, orbital entanglement allows us to monitor the change in correlation effects when chemical bonds are stretched. For instance, the dissociation process of the N$_2$, CsH, and TlH molecules reveals that these systems can be well-represented
by a single-reference approach around the equilibrium bond length, while a multi-reference treatment is required for large inter-nuclear separations.~\cite{entanglement_bonding_2013,Knecht2014}
Furthermore, orbital entanglement studies on iron nitrosyl complexes~\cite{entanglement_letter} demonstrate the effect of the $3d'$-shell on orbital entanglement and correlation. It is well-known that the addition of a $3d'$-shell is essential for a qualitative correct description of correlation effects within the $3d$-shell. Orbital entanglement thus provides an alternative understanding of the importance of the so-called ``double $d$-shell''-effect for transition metal systems in active space calculations.

\begin{figure*}[bt]
\caption{\label{fig:ee}
Orbital-pair mutual information and single-orbital entropy for stretched H$_2$ ($d_{\rm H-H}=2.0$ \AA{}, cc-pVTZ) determined from the antisymmetric product of 1-reference orbital geminals.~\cite{OO-AP1roG,ap1rog-jctc} The orbitals labelled by indices (1) and (2) correspond to the $\sigma$- and $\sigma^*$-orbitals.
}
\vspace*{-0.5cm}
\begin{center}
\includegraphics[width=0.99\textwidth,keepaspectratio=true]{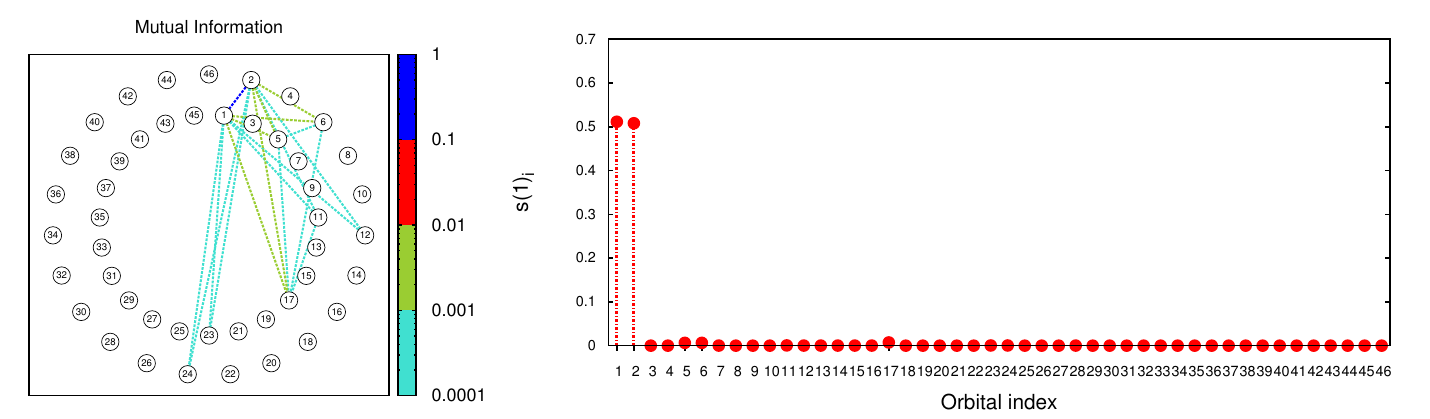}
\end{center}
\vspace*{-0.5cm}
\end{figure*}

\subsection{How to Choose the Active Space?}
The selection of the active space is a crucial step in multi-configuration self-consistent field calculations. Specifically, the active space should contain all orbitals that are required to describe strong electron correlation and to obtain a reliable zeroth-order wavefunction. A common procedure for examining the stability of active spaces is to study different sizes of the active space and analyse the convergence of the energy or related properties with respect to its size. Orbital entanglement can be particularly instructive in choosing the active space orbitals and in detecting unbalanced active spaces. If the active space is not chosen appropriately, strong electron correlation effects are overestimated, \textit{i.e.}, orbitals are entangled too strongly.~\cite{entanglement_letter,entanglement_bonding_2013}
Comparing orbital entanglement/correlation diagrams for different active spaces allows us to assess the quality, convergence behaviour, and possible artefacts of (too small) active space calculations.

\subsection{Chemical Bonding}
Since strong (nondynamic) electron correlation is important to allow a molecule to correctly dissociate into its fragments, orbital entanglement offers a conceptual understanding of bond-forming and bond-breaking processes~\cite{entanglement_bonding_2013} and can be used to gain insights into chemical processes. The bonding and anti-bonding orbital pairs that constitute ionic and covalent bonds in molecules become highly correlated at large inter-nuclear separation (see Figure~\ref{fig:ee} as an example). Specifically, those orbitals that are involved in bonding exhibit large values of $s(1)_i$ near the dissociation limit, while a bond is considered to be broken when $s(1)_i$ reaches its maximum value of $\ln 4$~\cite{Rissler_2006,entanglement_bonding_2013} (for spatial orbitals). 
For example, orbital entanglement correctly predicts single bonds in the F$_2$ and CsH molecules as well as s single C-C bond in ethane, a double bond in ethene, and a triple bond in the N$_2$, HCP and [CP$^-$] molecules as well as triple C-C bond in acetylene.~\cite{entanglement_bonding_2013,PCCP_bonding}

Furthermore, the rate of growth in $s(1)_i$ allows us to qualitatively resolve bond-breaking
processes of individual $\sigma$-, $\pi$-, etc., bonds in multiple bonding.~\cite{entanglement_bonding_2013,PCCP_bonding}
We observed that the rate of growth in $s(1)_i$ depends on the type (or entanglement strength) of a specific bond
as the one- and two-orbital entanglement measures are calculated from the electronic wave
function. For example, single orbital entropies of orbitals involved in weak (shorter) $\pi$-bonds
in the N$_2$ and HCP molecules increase faster than those corresponding to strong $\sigma$-bonds.

Our recent study on the chemical reaction pathway of nickel-ethene complexation~\cite{Corinne-2014} demonstrates that orbital entanglement can be used to monitor the evolution of bond formation processes even in cases where the orbital picture completely fails. Orbital entanglement confirms that metal-ligand bonding is initiated by metal-to-ethene back-donation, which begins around the transition state. This back-donation is followed by $\pi$-donation from the ethene ligand to the metal center, which leads to a metal-ligand bond.
This reaction mechanism is in perfect agreement with well-established metal-olefin bond models. Orbital entanglement can thus be considered as an alternative to molecular orbital theory, providing quantitative means to assess orbital interactions.
Most importantly, an orbital entanglement analysis can be applied even in cases where the simple picture of interacting orbitals fails, as in the case of nickel-ethene complexation.

Concepts of quantum information theory are also instrumental in unravelling the effect of noble gas coordination on the ground state of the CUO molecule.~\cite{CUO_DMRG}
The decay of the orbital-pair mutual information in the equatorially coordinated CUONe$_4$ and CUOAr$_4$ compounds indicated that the valence orbitals of Ar$_4$ are more correlated than those of Ne$_4$ with molecular orbitals centered on the CUO unit.~\cite{CUO_DMRG}
This observation agrees with experiment,~\cite{CUO_Nb_science} and elucidates the stabilization of the triplet state of
CUOAr$_4$ compared to its singlet state.~\cite{CUO_DMRG}

\subsection{Changes in the Wavefunction Character}
Concepts from quantum information theory can also be used to detect points where electronic wavefunctions change drastically.
Legeza and co-workers~\cite{Ors-LiF-TTNS} showed that the total quantum information ($I_{\rm tot}$) exhibits a discontinuity at the ionic-neutral curve-crossing point of the LiF molecule. A similar observation in the evolution of the total quantum information was made in the nickel-ethene complexation pathway where $I_{\rm tot}$ had a maximum in the transition state. Recently, Fertitta \textit{et al.}~\cite{Beate-Ors-Be-rings} used $I_{\rm tot}$ to successfully locate the position of the metal-insulator transition in Be rings.
In all cases, (avoided curve crossing, transition state, and metal-insulator transition) the electronic wavefunction changes considerably.

\section{Conclusions and Outlook\\}
The picture of interacting orbitals is an important concept in chemistry. It is used to understand the mechanisms of chemical reactions, chemical bonding, and ligand-metal interactions. Conventional approaches that assess the interaction of orbitals are, however, limited to a qualitative picture. Concepts from quantum information theory can provide quantitative means to analyse orbital interactions. Specifically, the entanglement entropy of orbitals and the orbital-pair mutual information allow us to quantify the entanglement and correlation of orbitals and pairs of orbitals.
Orbital entanglement and correlation measures can be used to dissect electron correlation effects into different contributions, to analyse the stability and the convergence of active space calculations, to gain insights into chemical reactions and mechanisms, and to identify points where electronic wavefunctions change dramatically (avoided crossings, transition states, metal-insulator transitions). If orbital entanglement is used to elucidate chemical processes, accurate electronic wavefunctions are required to determine the one- and two-orbital reduced density matrices. This, however, represents a major challenge in transition metal, lanthanide, and actinide chemistry, and may hamper the application of orbital entanglement to interesting domains of chemistry.

Despite all the aforementioned successes of orbital entanglement for interpreting and understanding chemical bonding, an entanglement analysis requires correlated wavefunctions as input. That means, orbital entanglement does not provide bonding information from one-determinant wavefunctions, like Hartree--Fock and Kohn--Sham density functional theory, as well as for the simple H$_2^+$ two-center-one-electron-bond model (because there is no electron correlation). Furthermore, since (entanglement-based) bond orders are deduced from changes in $s(1)_i$ and $I_{i|j}$, it is necessary to stretch the atomic centers of interest to observe these changes. This drawback hampers the application of orbital entanglement in predicting bond orders for large molecular systems, especially those that are far from a linear structure, \textit{e.g.}, C$_{60}$. Therefore, possible applications of orbital entanglement should be directed at dissecting electron correlation effects and analysing changes in electronic structure (transition states, curve crossings, etc.).

The one- and two-orbital reduced density matrices form the central ingredient of the orbital entanglement entropy and the orbital-pair mutual information. In contrast to entanglement measures based on $n$-particle RDMs, the one- and two-orbital reduced density matrices are determined from RDMs of different particle number. Their matrix elements can be determined from specific elements of the 1-, 2-, 3-, and 4-RDM. This property facilitates the determination of orbital entanglement and correlation using different quantum chemical methods, provided the $n$-particle RDMs are $N$-representable. Since response density matrices might not be $N$-representable, one should be cautious when interpreting orbital-entanglement measures for non-variational approaches, like coupled-cluster theory and similar approaches.~\cite{Tamar-pCC,ap1rog-jctc}

As the orbital entanglement entropy and the orbital-pair mutual information depend on the choice of the (molecular) orbital basis, future research of orbital entanglement should focus on studying the influence of the orbital basis on orbital entanglement and correlation and how the choice of the (molecular) orbital basis affects the interpretation of chemical processes. A first DMRG study in this direction was carried out by Fertitta \textit{et al.} who investigated the crucial dependence of entanglement on the (molecular) orbital basis and its severe impact on DMRG convergence.~\cite{Beate-Ors-Be-rings} However, deeper insight into the relation of orbital entanglement and orbital basis during chemical processes would be desirable.

Furthermore, since only a few elements of the 1-, 2-, 3-, and 4-RDMs are required to determine the one- and two-orbital RDMs, conventional quantum chemistry software packages can be easily extended to determine the orbital entanglement entropy and the orbital-pair mutual information. We believe that an orbital entanglement analysis of correlated electronic wavefunctions will be incorporated into standard quantum chemistry packages and complement conventional procedures for interpreting electronic structure and its changes, such as CI expansion coefficients, occupation numbers, and excitation amplitudes, in quantum chemical systems.

\section{Acknowledgement}
K.B. acknowledges the financial support from the Swiss National Science Foundation (P2EZP2 148650). P.T. gratefully acknowledges the financial support from the Natural Sciences and Engineering Research Council of Canada. We are grateful to Prof.~Paul W.~Ayers and Prof.~\"{O}rs Legeza for many helpful discussions.

\bibliography{rsc} 
\end{document}